\documentclass[aps,prb,groupedaddress,reprint,showpacs]{revtex4-1}
\usepackage{graphicx}
\usepackage{bm}
\usepackage{color}
\begin{document}
\title{First-Principles Study of Structural Transition in LiNiO$_2$ and
  High Throughput Screening for Long Life Battery}

\author{Tomohiro Yoshida$^{1}$}
\author{Kenta Hongo$^{2,3}$}
\author{Ryo Maezono$^{4,5}$}

\affiliation{$^{1}$
  Department of Computer-Aided Engineering and Development,
  Sumitomo Metal Mining Co., Ltd., Minato-ku, Tokyo 105-8716, Japan
}

\affiliation{$^{2}$
  Research Center for Advanced Computing Infrastructure,
  JAIST, Asahidai 1-1, Nomi, Ishikawa 923-1292, Japan
}

\affiliation{$^{3}$
  PRESTO, Japan Science and Technology Agency, 4-1-8 Honcho,
  Kawaguchi-shi, Saitama 322-0012, Japan
}

\affiliation{$^{4}$
  School of Information Science, JAIST, Asahidai 1-1, Nomi,
  Ishikawa, 923-1292, Japan
}

\affiliation{$^{5}$
  Computational Engineering Applications Unit, RIKEN, 2-1 Hirosawa,
  Wako, Saitama 351-0198, Japan
}

\date{\today}

\begin{abstract}
Herein, we performed {\it ab initio} screening to 
identify the best doping of LiNiO$_2$ to achieve 
improved cycle performance in lithium ion batteries. 
The interlayer interaction that dominates the $c$-axis 
contraction and overall performance 
was captured well by density functional theory 
using van der Waals exchange-correlation functionals. 
The screening indicated that Nb-doping is promising for improving cycle performance. 
To extract qualitative reasonings, 
we performed data analysis 
in a materials informatics manner to obtain
a reasonable regression to reproduce the obtained results. 
LASSO analysis implied that the charge density between 
the layers in the discharged state is the dominant
factor influencing cycle performance. 
\end{abstract}
\maketitle
\section{INTRODUCTION}
\label{sec.intro}
Improvements in lithium ion batteries (LIBs) have 
contributed to 'clean energy' as 
a key technology to reduce energy consumption and have been used in 
mobile devices and automobiles~\cite{Kang,Manthiram}. 
Cathode materials of LIBs most significantly influence
battery performances, especially 
in terms of their lifetime and safety (strength 
against damage and deteriorations)~\cite{Yoon,Gao,Dahn2}. 
Significant efforts~\cite{Nitta} have been made to
improve these aspects 
beyond investigations of the conventional cathode material, 
LiCoO$_2$~(LCO)~\cite{Mizushima}. 

\vspace{2mm}
LiNiO$_2$~(LNO) has been shown to achieve higher
energy densities during storage than that of LCO~\cite{Myung,Dahn,Ohzuku,Yamada,Rougier,Arai,Lee,Kanno,Li,Yoon}.  
It has also attracted interest because Ni is cheaper than Co.
The drawback of LNO is its cycle characteristics, 
which are relevant to the batteries' lifetime. 
The battery storage capacity is
reduced with increasing numbers of 
charge-discharge cycles.
After a smaller number of cycles, 
LNO loses its initial capacity more rapidly than the other cathode 
materials~\cite{Myung,Dahn,Ohzuku,Yamada,Rougier,Arai,Lee,Kanno,Li,Yoon}. 
Atomic substitution has been reported 
as a promising strategy to improve
cycle characteristics, including substituting Ni sites for
Al~\cite{Guilmarda,Guilmardb}, Mg~\cite{Pouillerie,Sathiyamoorthi,Kondo}, 
Co~\cite{Cho, Delmas, Saadoune}, 
Mn~\cite{Rossen,Venkatraman}, Fe~\cite{Mohana}, Y~\cite{Mohanb}, 
Ti~\cite{Kwon}, Zr~\cite{Kim,Yoona,Yoonb}, and Na~\cite{Kimb}. 

\vspace{2mm}
When $x$ percent of Li ions are removed from the cathode 
during charging, the cathode undergoes a 
structural transition from rhombohedral~(H1) [$x<0.25$] to 
monoclinic~(M) [$0.25<x<0.55$], rhombohedral~(H2) [$0.55<x<0.75$], 
and rhombohedral~(H3) [$0.75<x$]~\cite{Ohzuku,Yoon,Amatucci} successively. 
Drastic contractions along the $c$-axis occur when H2 transforms into H3, 
which has been identified as the major process causing deterioration 
during charge-discharge cycles~\cite{Yoon}. 
These contractions accumulate internal stresses, 
leading to cathode cracking and allowing 
the electrolyte to enter and cause further deterioration. 

\vspace{2mm}
Differences in cycle performances between LNO and LCO 
can be attributed to the different values of $x$ 
at the specific charging voltage, $4.2$~V. 
This voltage is the upper bound where 
the highest charging performance is realized while preventing 
electrolyte degradation. 
At the voltage (4.2 V), $x$ approaches $1.0$ 
the H2-H3 transition inevitably occurs in 
LNO~\cite{Ohzuku,Yoon}, 
whereas $x$ can be kept smaller to avoid this transition 
in LCO~\cite{Amatucci}. 
The cycle performance of LNO has been 
improved by reducing the 
charging voltage~\cite{Yoon} via reduced $x$ 
to hinder the large contractions along the $c$-axis. 
Nevertheless, achieving higher voltages is necessary to
improve charging capacities 
while suppressing the $c$-axis contraction 
by tuning LNO. 

\vspace{2mm}
Therefore, 
because the above described contraction dominates 
cycle performance, 
the computational design of the optimum doping
is necessary to realize improved 
performance as the contraction can be easily
evaluated by {\it ab initio} methods. 
A similar study was reported for 
LiFePO$_4$-based batteries, achieving 
drastic performance improvements ~\cite{Nishijima}. 
Herein, we performed theoretical calculation to 
identify which doping element could result in the best cycle 
performance. 
The $c$-axis contractions were evaluated by 
{\it ab initio} methods with 32 candidate elements 
as the dopant 
to substitute the Ni sites in LNO. 
This high-throughput screening  
showed that the Nb-doping
would result in the best performance. 

\section{METHODS}
\label{sec.fpc}
\begin{figure}[tb]
  \includegraphics[width=\hsize]{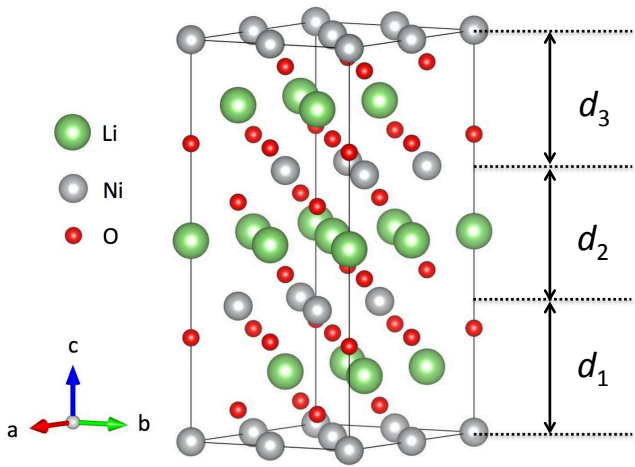}
  \caption{The unit cell used to model the systems with dopants, 
    with a rhombohedral~(R-3m) symmetry. 
    The cell corresponds to the $2\times 2\times 1$ 
    supercell of the pristine LiNiO$_2$ unit cell. 
Inter-layer distances, $d_{i=1\sim 3}$, 
were averaged for evaluation 
as a measure for composition optimization.
}
  \label{fig.fig1}
\end{figure}

Fig.~\ref{fig.fig1} shows the unit cell we used 
to model the systems with dopants 
with a rhombohedral~(R-3m) symmetry~\cite{Ohzuku}. 
The cell corresponds to the $2\times 2\times 1$ 
supercell of the pristine LiNiO$_2$ unit cell. 
The Vienna Ab initio Simulation Package (VASP)~\cite{Kressea,Kresseb} 
was used for all density functional theory (DFT) evaluations 
with projector augmented wave (PAW) treatment 
for ionic cores. 
Using careful convergence tests, 
we used the plane wave cutoff ($E_{\rm cut}$) 
of $650$~eV with a $5\times 5\times 2$ $k$-mesh.

\vspace{2mm}
For the exchange-correlation (XC) functional, 
we adopted optB86b-vdW~\cite{Klimesa,Klimesb} 
because conventional XC such as PBE~\cite{Predew1} 
cannot capture the long-range binding 
between Ni-O layers,
especially after Li ions are removed during 
charging. 
Several choices are available for the van der Waals exchange-correlation functionals (vdW-XC). 
For LNO, the vdW-D3~\cite{Grimme} has been reported to
be incapable of reproducing the contraction 
along the $c$-axis~\cite{Chakraborty}. 
For LCO~\cite{Aykol} it has been reported
that the 'opt-' vdW-type XC 
well reproduce the unit cell volumes 
and voltages where Li ions begin desorbing, 
as estimated by Eq.~(\ref{eq.eq1}). 
The optB86b-vdW-XC~\cite{Klimesa,Klimesb} was 
confirmed to achieve the best performance 
to reproduce experimental results of many 
systems including metallic, ionic, and 
covalent crystals~\cite{Klimesa}. 
As described later, 
we confirmed that the vdW-XC is indispensable 
for describing the system studied here. 

\section{RESULTS AND DISCUSSION}
\begin{figure*}[tb]
  \begin{center}
    \begin{tabular}{c}
      \begin{minipage}{0.5\hsize}
        \includegraphics[width=75mm]{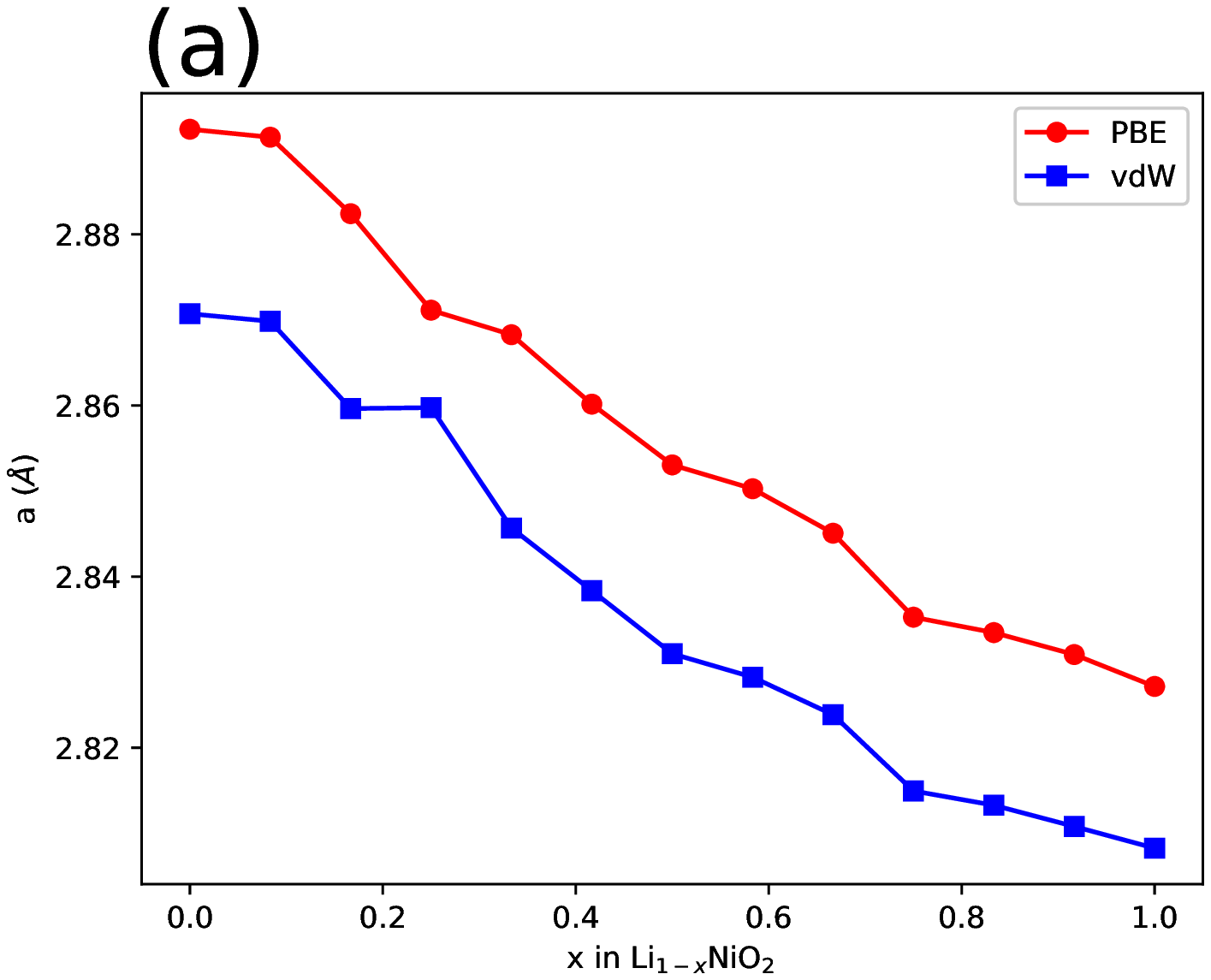}
      \end{minipage}
      \begin{minipage}{0.5\hsize}
        \includegraphics[width=75mm]{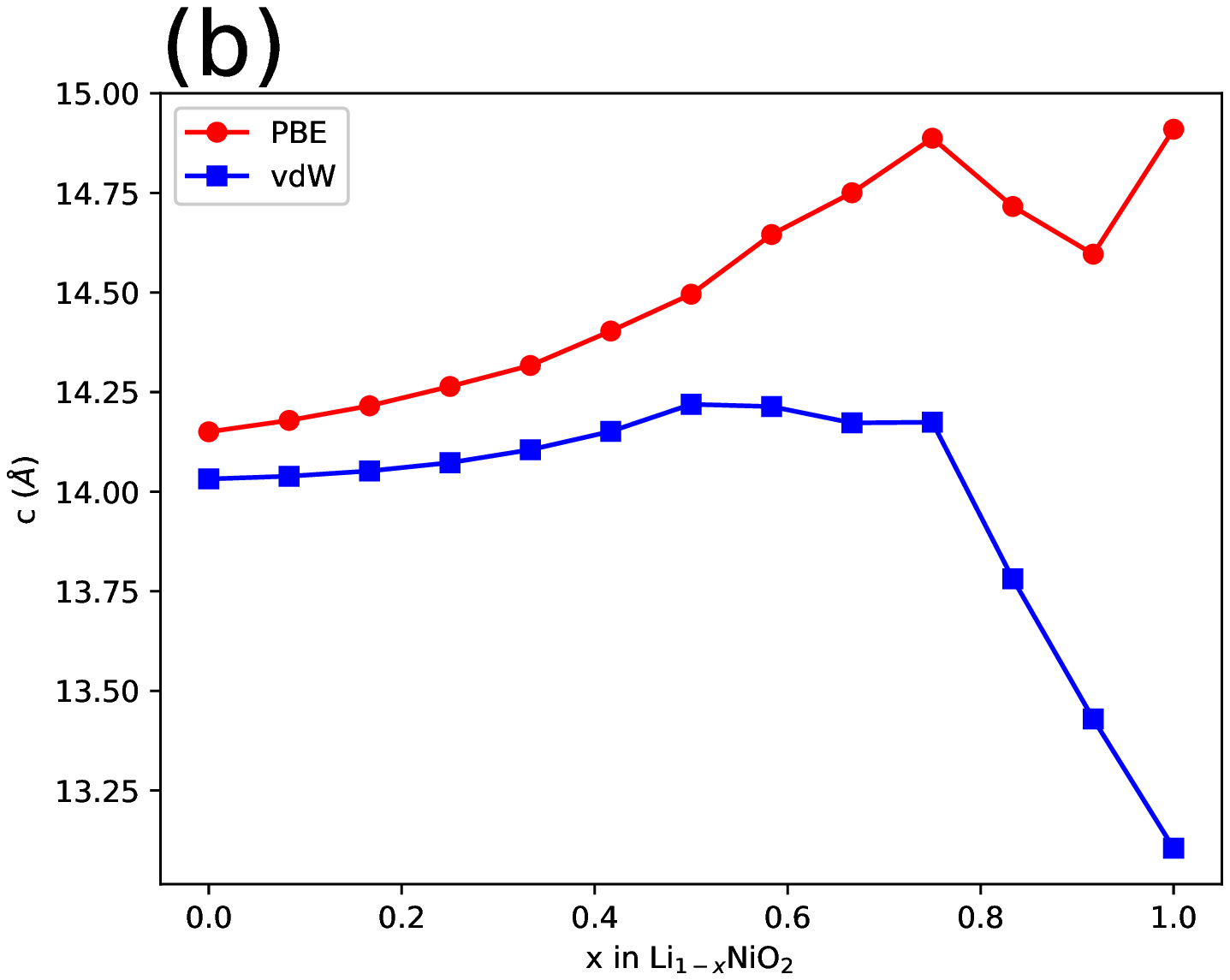}
      \end{minipage}
    \end{tabular}
    \caption{
Estimated lattice relaxation during the 
charge ($x=1$)-discharge ($x=0$) process with a 
comparison between PBE and vdW-XC. 
PBE was incapable of
reproducing the contraction along 
the $c$-axis during charging. 
}
    \label{fig.fig2}
  \end{center}
\end{figure*}

\vspace{2mm}
Table~\ref{table.table1} summarizes the optimized geometries 
and voltage at which the Li ions begin desorbing, $V_{\rm des}$, 
estimated using different XC and compared with 
experimental values~\cite{Ohzuku,Tamura}. 
The desorbing voltage
\[
  {\rm Li}_{x_1}{\rm NiO}_2
  \rightarrow {\rm Li}_{x_2}{\rm NiO}_2 + (x_1 - x_2)\cdot {\rm Li} \ ,
\]
can be estimated by 
\begin{eqnarray}
  V(x_1,x_2) = -\frac{E_{\rm Li_{x_1} NiO_2}
  -E_{\rm Li_{x_2} NiO_2}-(x_1-x_2)\cdot 
  E_{\rm Li}}{(x_1-x_2)\cdot e}, \nonumber \\ 
\label{eq.eq1}
\end{eqnarray}
where $E$ denotes the ground state energy of
each compound.
In Table~\ref{table.table1} we defined 
$V_{\rm des}=V(1,0)$. 
It is clear that vdW-XC achieved better
agreement with the experimental values 
than PBE for the geometry 
[(2\% by vdW) vs (11\% by PBE), 
error in $c$  the charged state (NiO$_2$)], 
and voltage [23\% vs 27\%]. 
Interestingly, the DFT+$U$ scheme 
further improved the agreement when $U$ was introduced to Ni-3$d$.
Using $U=6.7$~eV~\cite{Zhou}, errors of
1\% and 4\% were achieved for $c$ and $V_{\rm des}$. 
However, the '+$U$' scheme was not adopted in this study, {\it i.e.}, high-throughput screening. 
The $c$-length was used as the assessment 
function for screening and '+$U$' was only improved by a 
negligible amount compared to the magnitude of interest 
({\it i.e.}, the variation of $c$ during charging). 
As in Fig.~\ref{fig.fig2}, the trend 
required for the assessment function was well
captured even at $U=0$.  
The most compelling reason for not including '+$U$' is 
its high computational cost and its significantly worse convergence in a
SCF~(self-consistent field), rendering the screening 
inefficient. 
Furthermore, it is difficult to choose a proper
$U$ value for each system with different 
concentrations of Li vacancies and various possible 
atomic substitutions. 
\begin{table*}[htb]
  \begin{tabular}{lccccc} 
    \hline
    XC & $V_{\rm des}$ & $a$ (\AA) (LiNiO$_2$) & $c$(\AA) (LiNiO$_2$) 
    & $a$ (\AA) (NiO$_2$) & $c$(\AA) (NiO$_2$) \\
    \hline
    PBE & 3.05 & 2.89 & 14.16 & 2.82 & 14.85\\
    vdW & 3.25 & 2.87 & 14.04 & 2.81 & 13.10 \\
    Experiments\cite{Ohzuku,Tamura}  & 4.2 & 2.88 & 14.18 & 2.81 & 13.36 \\
    \hline
  \end{tabular}
  \caption{
  Predictions depending on the choice of 
  exchange-correlation functionals (XC). 
  $V_{\rm des}$ denotes the Li-desorbing voltage, 
  as defined in Eq.~(\ref{eq.eq1}) with the 
  optimized lattice parameters ($a$ and $c$) 
  along the $a$- and $c$-axes.  
  }
  \label{table.table1}
\end{table*}

\vspace{2mm}
Fig.~\ref{fig.fig2} shows the estimated 
lattice relaxations during the charge ($x=1$)-discharge ($x=0$)
process obtained using PBE and vdW-XC. 
For each $x$, relaxation calculations were performed
over all possible symmetries with the Li vacancies 
in the unit cell (shown in Fig.~\ref{fig.fig1}). 
The plot shows the lattice parameters of the most 
stable structure at each $x$ value. 
For the $a$-axis [Fig.~\ref{fig.fig2}(a)], 
both PBE and vdW-XC reproduced the trend where 
the lattice parameter was reduced monotonically  
as the system is charged, consistent
with experimental observations. 
A remarkable contrast was observed in the $c$-axis 
[Fig.~\ref{fig.fig2}(b)], where PBE failed to reproduce 
the contraction as $x$ increases. 
Up to $x=0.8$, it was confirmed that 
at least one Li ion present within 
the inter-layers prevented contraction, 
but at $x=0.8$ this inter-layer appears without 
Li, allowing for sudden contractions. 
The increasing trend for $0<x<0.8$ was attributed to 
the reduced ionic interactions caused by the charge
reduction in the Ni-O layer by 
the oxidation of Ni$^{3+}$ to Ni$^{4+}$. 
The large overestimations of the PBE lattice constants
can be primarily attributed to the general 
tendency of PBE to afford underbindings~\cite{2012OUM}, 
and to the lack of vdW inter-layer attractions 
that only the vdW-XC considered.
Since the contribution of the vdW forces to the cohesion was 
estimated to be approximately $1\sim 2\%$~\cite{Kittel}, 
the difference in the plot (Fig.~\ref{fig.fig2}) 
of approximately 1$\%$ is reasonable. 
Even with reduced ionic binding between layers, 
PBE yielded contractions at $0.8<x<0.9$ 
to fill the Li vacancies between the layers. However, 
at $x=1.0$ the reduced charge 
eventually becomes incapable of forming ionic 
bonds between layers~\cite{Aykol}, leading to increased 
inter-layer spacing. On the other hand,
using vdW-XC, the binding works 
even with the reduced charge, reproducing the
proper contractions to fill Li vacancies. 

\vspace{2mm}
The sharp decrease in the $c$ parameter [the blue plot in Fig.~\ref{fig.fig2}(b)] 
corresponds to the H2$\rightarrow$H3 transition~\cite{Ohzuku,Yoon}. 
It should be noted that this change is not accompanied by
any symmetric transitions, so it is not a true 'structural transition' but 
a 'structural change'. 
The main conclusion is that vdW-XC is indispensable for describing the H2$\rightarrow$H3 process. 
However, previous studies~\cite{Chakraborty} successfully
employed meta-GGA to describe this process. 
Although meta-GGA does not include dispersion interactions 
explicitly, it has been reported~\cite{Hongo} that 
it can describe vdW-like $\sim r^{-6}$ behavior 
in the system.
We employed vdW-XC rather than meta-GGA 
because the latter gives rise to worse SCF convergence~\cite{Yao},
which is fatal for high-throughput screening. 

Confident of our XC choice 
of vdW/$U\!=\!0$ describing the target,
namely the $c$-axis contractions as an indicator of cycle performance, 
we performed high-throughput screening to determine
the optimum choice of the doping element
for LNO. 
We considered 66 doping elements, 
'X', from the third row and 
lower in the periodic table, up to Pa 
(atomic number 91) and excluding those 
for which pseudo potentials were not available. 
To enhance rate capability, Co is usually doped into LNO~\cite{Schipper01012017}.
We also substituted X to Ni to improve cycle performance.
The ratio of elements was Ni:Co:X=0.75:0.17:0.08, 
{\it i.e.}, where two Ni sites are occupied by Co atoms, 
while one Ni site is occupied by X within a unit cell, as depicted 
in Fig.~\ref{fig.fig1}.
It should be noted that some ions cannot adopt a trivalent state. 
For example, Na only adopts a stable valence of Na$^{+1}$. 
However, to ensure charge neutrality in such cases, the valence of two Ni atoms can be changed from Ni$^{+3}$ to Ni$^{+4}$. 
Therefore, it was assumed that all elements can be dissolved in LNO.
It should be mentioned that the supercell size ($2\times2\times 1$) was relatively small.
However, if a larger cell was considered, significant amounts of time would be necessary to determine the most stable structure.
Therefore, we used the $2\times2\times 1$ supercell and 
will investigate larger systems using genetic algorithms or Bayesian optimization in the future.
There are $_{12}{\rm C}_2 \times 10 = 660$ possible substitution patterns that can be achieved under the above conditions, 
but we can reduce these structures to 10 symmetrically different structures 
(see Appendix A).
For these 10 possibilities,
geometrical optimizations were performed to determine the most 
stable structure as a representative for X 
to be compared with the structures obtained from other X. 
We compared the $c$-axis contraction, 
$\Delta d_{\rm ave}=\sum_{i=1}^{3} (d_i^d - d_i^c)/3$, 
where $d_i^d$ ($d_i^c$) denotes the layer
spacing, as depicted 
in Fig.~\ref{fig.fig1} when the system is 
discharged (charged). 

The spacing, $d_i^c$, of the charged state 
was evaluated at the $75~\%$-charged state, 
{\it i.e.}, the crystal structure where 
9 Li ions are removed from the unit cell 
shown in Fig.~\ref{fig.fig1}. 
This percentage was the maximum 
where we could expect to maintain the charge 
neutrality because only Ni$^{3+}$ to Ni$^{4+}$ and Co$^{3+}$ to Co$^{4+}$
could compensate for neutrality, while other ions 
could not due to their specific redox potentials. 
For the charged states, geometry optimizations 
were performed to determine $d_i^c$ starting from 
the initial structures generated by LNO with 
Li vacancies.

It is well-known that the ground state of LNO is ferromagnetic~\cite{PhysRevB.44.943}, 
but when Ni is substituted by other ions, the ground state magnetic structure changes.
However, as shown in Appendix B, the geometry was hardly changed by magnetic order, 
so calculations were performed with ferromagnetic spin polarization.

\begin{figure*}[tb]
  \includegraphics[width=\hsize]{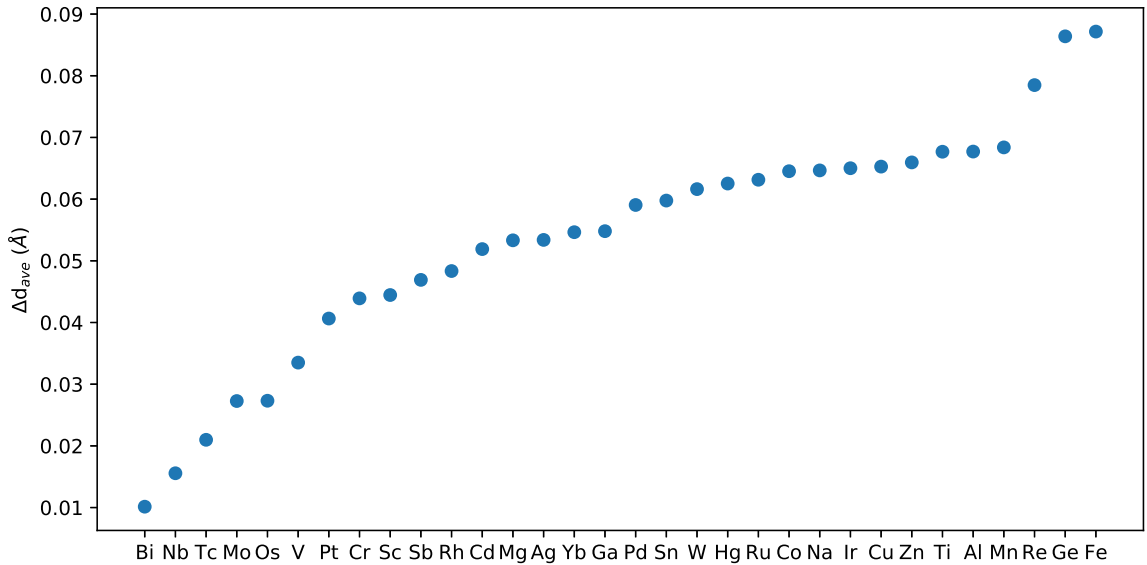}
  \caption{The $c$-axis contractions induced by 75\% charging were 
evaluated in terms of $\Delta d_{\rm ave}$. 
Lower vertical axis values indicate 
smaller expected contractions and
improved cycle performance. 
Elements along the horizontal axis are arranged 
in order of the magnitude of $\Delta d_{\rm ave}$. 
}
  \label{fig.fig3}
\end{figure*}

\vspace{2mm}
Fig.~\ref{fig.fig3} summarizes the screening results. 
Decreasing vertical axis values indicates 
smaller contractions and improved 
cycle performance.
Cases that
yielded larger structural variation with 
axes ($a$, $b$, and $c$) tilted more than 2$^\circ$
during charging were excluded from the plot[Sr, Zr, Eu, Dy, Ho, Er, Tm, Lu, Pb, K, 
Rb, La, Ra, Au, Ce, Ca, Ba, Ta, Y, Pm]. 
Negative $\Delta d_{\rm ave}$ values were also excluded
because the negative values
indicate that the contraction is already 
larger than that of LNO (positive at 75\% charging). 
Among the 32 elements depicted, 
Bi and Nb doping are 
expected to yield the smallest contractions and most significantly improve cycle performance.
To investigate thermodynamic stability, 
the formation energies of the Bi- and Nb-doped systems were calculated.
Bi has a high formation energy, so it is expected that Bi cannot be dissolved in LNO.
Therefore, we conclude that the best doping element is Nb.
(See Appendix C for further details.)

Although it is generally difficult to identify 
the factors underlying the trends observed in Fig.~\ref{fig.fig3}, 
we performed data analyses using
materials informatics, as described below. 
For the descriptors of $\Delta d_{\rm ave}$, 
we used 'elemental information of the X substitutes'
(including atomic numbers, atomic radii {\it etc.}) 
provided on the horizontal axis of Fig.~\ref{fig.fig3}.
The elemental information was obtained from 
'pymatgen'~(Python Materials Genomics)~\cite{Ong} 
to obtain a set of properties, as listed in the left column 
of Table~\ref{table.table2}. 
For X $=$ Hg and X $=$ Tc, incomplete 
data was available and they were
excluded from the regression. 
As descriptors, we included 
the 'structural information' (including substituted position 
of Co, which is another substitute in the system) 
as listed in the right column. 
\begin{table}[tb]
  \begin{tabular}{cc} 
    \hline
    Elemental information of X & Geometries at discharged state \\
    \hline
    Atomic number & Lattice constant \\
    Group &  Lattice volume \\
    Period & Substituted position of Co\\
    Covalent radius & Substituted position of X\\
    Electronegativity & \\
    First-ionization energy & \\
    Atomic mass & \\
    Atomic radius & \\
    Coefficient of linear thermal expansion & \\
    Solid density & \\
    Electrical resistivity & \\
    Molar volume & \\
    Thermal conductivity & \\
    \hline
  \end{tabular}
  \caption{
    List of descriptors for the regression of $\Delta d_{\rm ave}$. 
'X' denotes the element appearing 
on the horizontal axis of Fig.~\ref{fig.fig3}. 
The left-hand column lists the properties of X, 
which were obtained from the 
'pymatgen' database~\cite{Ong}. 
  }  
  \label{table.table2}
\end{table}

\vspace{2mm}
Less important descriptors were 
identified by LASSO 
regression~\cite{Tibshirani}. 
By excluding these descriptors, we obtained a sparse form
of a regression, 
\begin{eqnarray}
  \Delta d_{{\rm X},{\rm ave}} &=&
  -8.5\times 10^{-3}\cdot (x \ {\rm coordinate\ of \ X}) \nonumber \\
  && -6.8 \times 10^{-3}\cdot (z \ {\rm coordinate\  of \ Co}) \nonumber \\
  && -2.1 \times 10^{-3}\cdot ({\rm covalent \ radius}) \nonumber \\
  && +1.3 \times 10^{-3}\cdot (x \ {\rm coordinate\  of \ Co}) \nonumber \\
  && +7.8 \times 10^{-4}\cdot ({\rm density \ of \ solid}) \nonumber \\
  && -4.4 \times 10^{-4}\cdot c \nonumber \\
  && -8.6 \times 10^{-5}\cdot ({\rm atomic \ radius}) \ , 
  \label{eq.eq2}
\end{eqnarray}
where each descriptor is standardized
so that the standardized data set has a mean of 0.0 and standard deviation of 1.0. 
\begin{figure}[tb]
  \includegraphics[width=\hsize]{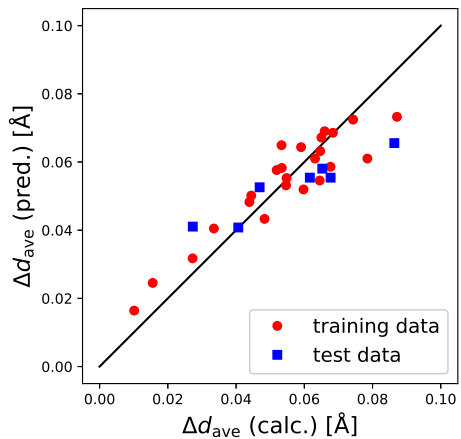}
  \caption{
The performance of the regression given in Eq.~(\ref{eq.eq2}). 
The regression (vertical axis) reproduced 
the simulation results (horizontal axis) fairly well
with an RMS error of $8.3\times10^{-3}$\AA.
}
  \label{fig.fig4}
\end{figure}
The regression, Eq.~(\ref{eq.eq2}), was confirmed 
to work fairly well, as shown in Fig.~\ref{fig.fig4}. 
It (vertical axis) reproduced the results obtained by the simulation 
(horizontal axis) within an RMS error of $8.3\times10^{-3}$\AA.

\vspace{2mm}
From the LASSO regression, we identified 
the quantities that should be considered 
for designing systems with improved cycle performance. 
The relevant factors are the substituted positions 
and densities of X or Co, as they are likely to dominate 
the charge density at the inter-layer region and contraction of the crystals via inter-layer vdW interactions. 
Because the regression was based only on the relaxed geometries 
and densities in the discharged state, 
we avoided the need for full-simulations including 
evaluations of the charged state, 
where it is difficult to consider Li vacancies. 
Once we are confident in
the qualitative observations that 
the charge density between layers in the 
discharged state was significant, 
we will attempt to search for 
complex compositions (such as
substitutions using two elements {\it etc.}) 
to achieve further improved performance. 

\section{CONCLUSION}
The $c$-axis contractions of LNO when 
Li ions are desorbed were 
satisfactorily described by DFT with vdW-XC, 
but not with PBE. 
The optimized framework reproduced the H2$\rightarrow$H3 
'transition' as a sharp drop of the $c$ length 
without symmetric transition. 
Using this framework, we performed 
a high-throughput screening for
ideal combinations of doping substitutes
for Co in LNO to minimize the change in $c$ for improved
cycle performance. 
Nb was the most promising 
candidates in this study.
The computational screening can
save significant amounts of time required for 
the materials screening by experimental 
syntheses. 
Using LASSO data analysis, the 
regression implied that the charge density 
between layers was the 
dominant underlying factor for the contraction. 

\label{sec.conc}

\begin{acknowledgements}
T.~Y. would like to thank K.~Utimula, T.~Kosasa, K.~Ryoshi, T.~Toma, S.~Yoshio, 
and N.~Watanabe for their fruitful discussions and technical support. 
The computations in this work were performed using the 
facilities of the Research Center for Advanced Computing Infrastructure at JAIST. 
K.H. is grateful for financial support from a KAKENHI grant (17K17762), 
a Grant-in-Aid for Scientific Research on Innovative Areas (16H06439),
the FLAG-SHIP2020 project (MEXT for the computational resources, projects hp180206 and 
hp180175 at K-computer),
PRESTO (JPMJPR16NA) and the Materials research
by Information Integration Initiative (MI$^2$I) project of the
Support Program for Starting Up Innovation Hub from Japan Science
and Technology Agency (JST).
R.M. is grateful for financial support from MEXT-KAKENHI (project JP16KK0097), 
the FLAG-SHIP2020 project (MEXT for the computational resources, projects hp180206 and 
hp180175 at K-computer),
and the Air Force Office of Scientific Research (AFOSR-AOARD/FA2386-17-1-4049)
\end{acknowledgements}


\begin{appendix}
\section{Calculated structures}
\begin{figure}[hbt]
  \includegraphics[width=\hsize]{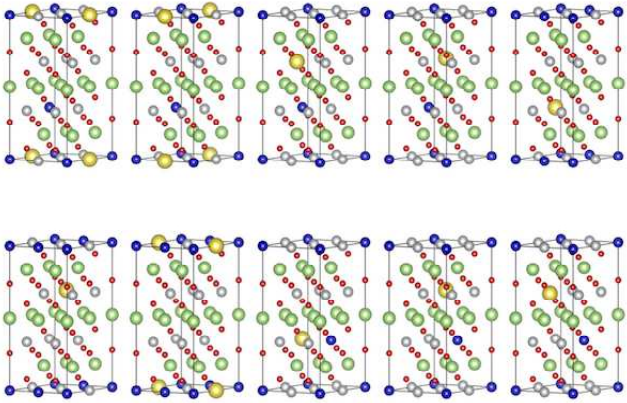}
  \caption{10 symmetrically different doped systems. Green, red, gray, blue, and yellow balls show Li, O, Ni, Co, and X, respectively}
  \label{fig.fig5}
\end{figure}

We show the caluculated strucures for doped systems in Fig.~\ref{fig.fig5}.

\section{Magnetic structure}
We sumarize energy and lattice parameters for ferromagnetic, antiferromagnetic stacking of spin-ferro orderd Ni layers, 
and non magnetic LNO in Table \ref{table.table3}.
\begin{table*}[hbt]
  \begin{tabular}{lccccc} 
    \hline
    Magnetic structure & Energy (meV) & $a$(\AA) & $c$(\AA) \\
    \hline
    Ferromagnetic order & 0 & 2.87 & 14.04\\
    Antiferromagnetic order & 19 & 2.88 & 14.06 \\
    Non magnetic order & 148 & 2.87 & 14.00 \\
    \hline
  \end{tabular}
  \caption{Relative energies and structural parameters. The zero for the energy is chosen to be ferromagnetic order.
  }
  \label{table.table3}
\end{table*}

\section{Formation energy}
Formation energy $E_{\rm F}$ is defined as,
\begin{eqnarray}
  E_{\rm F} &=& E({\rm LiNi_{0.75}Co_{0.17}X_{0.08}O}_2) \nonumber \\
  &&-(1-c)E({\rm LiNi_{0.83}Co_{0.17}O}_2)-cE({\rm LiX_{0.83}Co_{0.17}O}_2) \nonumber \\
\end{eqnarray}
where $E$ and $c=0.1$ denote the total energy and concentration of X.
We investigate the formation energy for Bi and Nb doped system.
The formation energy is 0.848 eV for former and -1.993 eV for later. 
\end{appendix}

\bibliography{references}

\begin{thebibliography}{53}%
\makeatletter
\providecommand \@ifxundefined [1]{%
 \@ifx{#1\undefined}
}%
\providecommand \@ifnum [1]{%
 \ifnum #1\expandafter \@firstoftwo
 \else \expandafter \@secondoftwo
 \fi
}%
\providecommand \@ifx [1]{%
 \ifx #1\expandafter \@firstoftwo
 \else \expandafter \@secondoftwo
 \fi
}%
\providecommand \natexlab [1]{#1}%
\providecommand \enquote  [1]{``#1''}%
\providecommand \bibnamefont  [1]{#1}%
\providecommand \bibfnamefont [1]{#1}%
\providecommand \citenamefont [1]{#1}%
\providecommand \href@noop [0]{\@secondoftwo}%
\providecommand \href [0]{\begingroup \@sanitize@url \@href}%
\providecommand \@href[1]{\@@startlink{#1}\@@href}%
\providecommand \@@href[1]{\endgroup#1\@@endlink}%
\providecommand \@sanitize@url [0]{\catcode `\\12\catcode `\$12\catcode
  `\&12\catcode `\#12\catcode `\^12\catcode `\_12\catcode `\%12\relax}%
\providecommand \@@startlink[1]{}%
\providecommand \@@endlink[0]{}%
\providecommand \url  [0]{\begingroup\@sanitize@url \@url }%
\providecommand \@url [1]{\endgroup\@href {#1}{\urlprefix }}%
\providecommand \urlprefix  [0]{URL }%
\providecommand \Eprint [0]{\href }%
\providecommand \doibase [0]{http://dx.doi.org/}%
\providecommand \selectlanguage [0]{\@gobble}%
\providecommand \bibinfo  [0]{\@secondoftwo}%
\providecommand \bibfield  [0]{\@secondoftwo}%
\providecommand \translation [1]{[#1]}%
\providecommand \BibitemOpen [0]{}%
\providecommand \bibitemStop [0]{}%
\providecommand \bibitemNoStop [0]{.\EOS\space}%
\providecommand \EOS [0]{\spacefactor3000\relax}%
\providecommand \BibitemShut  [1]{\csname bibitem#1\endcsname}%
\let\auto@bib@innerbib\@empty
\bibitem [{\citenamefont {Kang}\ \emph {et~al.}(2006)\citenamefont {Kang},
  \citenamefont {Meng}, \citenamefont {Br\'{e}ger}, \citenamefont {Grey},\ and\
  \citenamefont {Ceder}}]{Kang}%
  \BibitemOpen
  \bibfield  {author} {\bibinfo {author} {\bibfnamefont {K.}~\bibnamefont
  {Kang}}, \bibinfo {author} {\bibfnamefont {Y.~S.}\ \bibnamefont {Meng}},
  \bibinfo {author} {\bibfnamefont {J.}~\bibnamefont {Br\'{e}ger}}, \bibinfo
  {author} {\bibfnamefont {C.~P.}\ \bibnamefont {Grey}}, \ and\ \bibinfo
  {author} {\bibfnamefont {G.}~\bibnamefont {Ceder}},\ }\href@noop {}
  {\bibfield  {journal} {\bibinfo  {journal} {Science}\ }\textbf {\bibinfo
  {volume} {311}},\ \bibinfo {pages} {977} (\bibinfo {year}
  {2006})}\BibitemShut {NoStop}%
\bibitem [{\citenamefont {Manthiram}\ \emph {et~al.}(2008)\citenamefont
  {Manthiram}, \citenamefont {Vadivel~Murugan}, \citenamefont {Sarkar},\ and\
  \citenamefont {Muraliganth}}]{Manthiram}%
  \BibitemOpen
  \bibfield  {author} {\bibinfo {author} {\bibfnamefont {A.}~\bibnamefont
  {Manthiram}}, \bibinfo {author} {\bibfnamefont {A.}~\bibnamefont
  {Vadivel~Murugan}}, \bibinfo {author} {\bibfnamefont {A.}~\bibnamefont
  {Sarkar}}, \ and\ \bibinfo {author} {\bibfnamefont {T.}~\bibnamefont
  {Muraliganth}},\ }\href@noop {} {\bibfield  {journal} {\bibinfo  {journal}
  {Energy Environ. Sci.}\ }\textbf {\bibinfo {volume} {1}},\ \bibinfo {pages}
  {621} (\bibinfo {year} {2008})}\BibitemShut {NoStop}%
\bibitem [{\citenamefont {Yoon}\ \emph {et~al.}(2017)\citenamefont {Yoon},
  \citenamefont {Jun}, \citenamefont {Myung},\ and\ \citenamefont
  {Sun}}]{Yoon}%
  \BibitemOpen
  \bibfield  {author} {\bibinfo {author} {\bibfnamefont {C.~S.}\ \bibnamefont
  {Yoon}}, \bibinfo {author} {\bibfnamefont {D.-W.}\ \bibnamefont {Jun}},
  \bibinfo {author} {\bibfnamefont {S.-T.}\ \bibnamefont {Myung}}, \ and\
  \bibinfo {author} {\bibfnamefont {Y.-K.}\ \bibnamefont {Sun}},\ }\href@noop
  {} {\bibfield  {journal} {\bibinfo  {journal} {ACS Energy Lett.}\ }\textbf
  {\bibinfo {volume} {2}},\ \bibinfo {pages} {1150} (\bibinfo {year}
  {2017})}\BibitemShut {NoStop}%
\bibitem [{\citenamefont {Gao}\ \emph {et~al.}(1998)\citenamefont {Gao},
  \citenamefont {Yakovleva},\ and\ \citenamefont {Ebner}}]{Gao}%
  \BibitemOpen
  \bibfield  {author} {\bibinfo {author} {\bibfnamefont {Y.}~\bibnamefont
  {Gao}}, \bibinfo {author} {\bibfnamefont {M.~V.}\ \bibnamefont {Yakovleva}},
  \ and\ \bibinfo {author} {\bibfnamefont {W.~B.}\ \bibnamefont {Ebner}},\
  }\href@noop {} {\bibfield  {journal} {\bibinfo  {journal} {Electrochem.
  Solid-State Lett.}\ }\textbf {\bibinfo {volume} {1}},\ \bibinfo {pages} {117}
  (\bibinfo {year} {1998})}\BibitemShut {NoStop}%
\bibitem [{\citenamefont {Dahn}\ \emph {et~al.}(1994)\citenamefont {Dahn},
  \citenamefont {Fuller}, \citenamefont {Obrovac},\ and\ \citenamefont {von
  Sacken}}]{Dahn2}%
  \BibitemOpen
  \bibfield  {author} {\bibinfo {author} {\bibfnamefont {J.}~\bibnamefont
  {Dahn}}, \bibinfo {author} {\bibfnamefont {E.}~\bibnamefont {Fuller}},
  \bibinfo {author} {\bibfnamefont {M.}~\bibnamefont {Obrovac}}, \ and\
  \bibinfo {author} {\bibfnamefont {U.}~\bibnamefont {von Sacken}},\
  }\href@noop {} {\bibfield  {journal} {\bibinfo  {journal} {Solid State
  Ionics}\ }\textbf {\bibinfo {volume} {69}},\ \bibinfo {pages} {265} (\bibinfo
  {year} {1994})}\BibitemShut {NoStop}%
\bibitem [{\citenamefont {Nitta}\ \emph {et~al.}(2015)\citenamefont {Nitta},
  \citenamefont {Wu}, \citenamefont {Lee},\ and\ \citenamefont
  {Yushin}}]{Nitta}%
  \BibitemOpen
  \bibfield  {author} {\bibinfo {author} {\bibfnamefont {N.}~\bibnamefont
  {Nitta}}, \bibinfo {author} {\bibfnamefont {F.}~\bibnamefont {Wu}}, \bibinfo
  {author} {\bibfnamefont {J.~T.}\ \bibnamefont {Lee}}, \ and\ \bibinfo
  {author} {\bibfnamefont {G.}~\bibnamefont {Yushin}},\ }\href {\doibase
  https://doi.org/10.1016/j.mattod.2014.10.040} {\bibfield  {journal} {\bibinfo
   {journal} {Mat. Today}\ }\textbf {\bibinfo {volume} {18}},\ \bibinfo {pages}
  {252} (\bibinfo {year} {2015})}\BibitemShut {NoStop}%
\bibitem [{\citenamefont {Mizushima}\ \emph {et~al.}(1980)\citenamefont
  {Mizushima}, \citenamefont {Jones}, \citenamefont {Wiseman},\ and\
  \citenamefont {Goodenough}}]{Mizushima}%
  \BibitemOpen
  \bibfield  {author} {\bibinfo {author} {\bibfnamefont {K.}~\bibnamefont
  {Mizushima}}, \bibinfo {author} {\bibfnamefont {P.}~\bibnamefont {Jones}},
  \bibinfo {author} {\bibfnamefont {P.}~\bibnamefont {Wiseman}}, \ and\
  \bibinfo {author} {\bibfnamefont {J.}~\bibnamefont {Goodenough}},\
  }\href@noop {} {\bibfield  {journal} {\bibinfo  {journal} {Mater. Res.
  Bull.}\ }\textbf {\bibinfo {volume} {15}},\ \bibinfo {pages} {783} (\bibinfo
  {year} {1980})}\BibitemShut {NoStop}%
\bibitem [{\citenamefont {Myung}\ \emph {et~al.}(2017)\citenamefont {Myung},
  \citenamefont {Maglia}, \citenamefont {Park}, \citenamefont {Yoon},
  \citenamefont {Lamp}, \citenamefont {Kim},\ and\ \citenamefont
  {Sun}}]{Myung}%
  \BibitemOpen
  \bibfield  {author} {\bibinfo {author} {\bibfnamefont {S.-T.}\ \bibnamefont
  {Myung}}, \bibinfo {author} {\bibfnamefont {F.}~\bibnamefont {Maglia}},
  \bibinfo {author} {\bibfnamefont {K.-J.}\ \bibnamefont {Park}}, \bibinfo
  {author} {\bibfnamefont {C.~S.}\ \bibnamefont {Yoon}}, \bibinfo {author}
  {\bibfnamefont {P.}~\bibnamefont {Lamp}}, \bibinfo {author} {\bibfnamefont
  {S.-J.}\ \bibnamefont {Kim}}, \ and\ \bibinfo {author} {\bibfnamefont
  {Y.-K.}\ \bibnamefont {Sun}},\ }\href@noop {} {\bibfield  {journal} {\bibinfo
   {journal} {ACS Energy Lett.}\ }\textbf {\bibinfo {volume} {2}},\ \bibinfo
  {pages} {196} (\bibinfo {year} {2017})}\BibitemShut {NoStop}%
\bibitem [{\citenamefont {Dahn}\ \emph {et~al.}(1990)\citenamefont {Dahn},
  \citenamefont {von Sacken},\ and\ \citenamefont {Michal}}]{Dahn}%
  \BibitemOpen
  \bibfield  {author} {\bibinfo {author} {\bibfnamefont {J.}~\bibnamefont
  {Dahn}}, \bibinfo {author} {\bibfnamefont {U.}~\bibnamefont {von Sacken}}, \
  and\ \bibinfo {author} {\bibfnamefont {C.}~\bibnamefont {Michal}},\
  }\href@noop {} {\bibfield  {journal} {\bibinfo  {journal} {Solid State
  Ionics}\ }\textbf {\bibinfo {volume} {44}},\ \bibinfo {pages} {87} (\bibinfo
  {year} {1990})}\BibitemShut {NoStop}%
\bibitem [{\citenamefont {Ohzuku}\ \emph {et~al.}(1993)\citenamefont {Ohzuku},
  \citenamefont {Ueda},\ and\ \citenamefont {Nakayama}}]{Ohzuku}%
  \BibitemOpen
  \bibfield  {author} {\bibinfo {author} {\bibfnamefont {T.}~\bibnamefont
  {Ohzuku}}, \bibinfo {author} {\bibfnamefont {A.}~\bibnamefont {Ueda}}, \ and\
  \bibinfo {author} {\bibfnamefont {M.}~\bibnamefont {Nakayama}},\ }\href@noop
  {} {\bibfield  {journal} {\bibinfo  {journal} {J. Electrochem. Soc.}\
  }\textbf {\bibinfo {volume} {140}},\ \bibinfo {pages} {1862} (\bibinfo {year}
  {1993})}\BibitemShut {NoStop}%
\bibitem [{\citenamefont {Yamada}\ \emph {et~al.}(1995)\citenamefont {Yamada},
  \citenamefont {Fujiwara},\ and\ \citenamefont {Kanda}}]{Yamada}%
  \BibitemOpen
  \bibfield  {author} {\bibinfo {author} {\bibfnamefont {S.}~\bibnamefont
  {Yamada}}, \bibinfo {author} {\bibfnamefont {M.}~\bibnamefont {Fujiwara}}, \
  and\ \bibinfo {author} {\bibfnamefont {M.}~\bibnamefont {Kanda}},\
  }\href@noop {} {\bibfield  {journal} {\bibinfo  {journal} {J. Power Sources}\
  }\textbf {\bibinfo {volume} {54}},\ \bibinfo {pages} {209} (\bibinfo {year}
  {1995})}\BibitemShut {NoStop}%
\bibitem [{\citenamefont {Rougier}\ \emph {et~al.}(1996)\citenamefont
  {Rougier}, \citenamefont {Gravereau},\ and\ \citenamefont
  {Delmas}}]{Rougier}%
  \BibitemOpen
  \bibfield  {author} {\bibinfo {author} {\bibfnamefont {A.}~\bibnamefont
  {Rougier}}, \bibinfo {author} {\bibfnamefont {P.}~\bibnamefont {Gravereau}},
  \ and\ \bibinfo {author} {\bibfnamefont {C.}~\bibnamefont {Delmas}},\
  }\href@noop {} {\bibfield  {journal} {\bibinfo  {journal} {J. Electrochem.
  Soc.}\ }\textbf {\bibinfo {volume} {143}},\ \bibinfo {pages} {1168} (\bibinfo
  {year} {1996})}\BibitemShut {NoStop}%
\bibitem [{\citenamefont {Arai}\ \emph {et~al.}(1997)\citenamefont {Arai},
  \citenamefont {Okada}, \citenamefont {Sakurai},\ and\ \citenamefont {ichi
  Yamaki}}]{Arai}%
  \BibitemOpen
  \bibfield  {author} {\bibinfo {author} {\bibfnamefont {H.}~\bibnamefont
  {Arai}}, \bibinfo {author} {\bibfnamefont {S.}~\bibnamefont {Okada}},
  \bibinfo {author} {\bibfnamefont {Y.}~\bibnamefont {Sakurai}}, \ and\
  \bibinfo {author} {\bibfnamefont {J.}~\bibnamefont {ichi Yamaki}},\
  }\href@noop {} {\bibfield  {journal} {\bibinfo  {journal} {Solid State
  Ionics}\ }\textbf {\bibinfo {volume} {95}},\ \bibinfo {pages} {275} (\bibinfo
  {year} {1997})}\BibitemShut {NoStop}%
\bibitem [{\citenamefont {Lee}\ \emph {et~al.}(1999)\citenamefont {Lee},
  \citenamefont {Sun},\ and\ \citenamefont {Nahm}}]{Lee}%
  \BibitemOpen
  \bibfield  {author} {\bibinfo {author} {\bibfnamefont {Y.~S.}\ \bibnamefont
  {Lee}}, \bibinfo {author} {\bibfnamefont {Y.~K.}\ \bibnamefont {Sun}}, \ and\
  \bibinfo {author} {\bibfnamefont {K.~S.}\ \bibnamefont {Nahm}},\ }\href@noop
  {} {\bibfield  {journal} {\bibinfo  {journal} {Solid State Ionics}\ }\textbf
  {\bibinfo {volume} {118}},\ \bibinfo {pages} {159} (\bibinfo {year}
  {1999})}\BibitemShut {NoStop}%
\bibitem [{\citenamefont {Kanno}\ \emph {et~al.}(1994)\citenamefont {Kanno},
  \citenamefont {Kubo}, \citenamefont {Kawamoto}, \citenamefont {Kamiyama},
  \citenamefont {Izumi}, \citenamefont {Takeda},\ and\ \citenamefont
  {Takano}}]{Kanno}%
  \BibitemOpen
  \bibfield  {author} {\bibinfo {author} {\bibfnamefont {R.}~\bibnamefont
  {Kanno}}, \bibinfo {author} {\bibfnamefont {H.}~\bibnamefont {Kubo}},
  \bibinfo {author} {\bibfnamefont {Y.}~\bibnamefont {Kawamoto}}, \bibinfo
  {author} {\bibfnamefont {T.}~\bibnamefont {Kamiyama}}, \bibinfo {author}
  {\bibfnamefont {F.}~\bibnamefont {Izumi}}, \bibinfo {author} {\bibfnamefont
  {Y.}~\bibnamefont {Takeda}}, \ and\ \bibinfo {author} {\bibfnamefont
  {M.}~\bibnamefont {Takano}},\ }\href@noop {} {\bibfield  {journal} {\bibinfo
  {journal} {J. Solid State Chem.}\ }\textbf {\bibinfo {volume} {110}},\
  \bibinfo {pages} {216} (\bibinfo {year} {1994})}\BibitemShut {NoStop}%
\bibitem [{\citenamefont {Li}\ \emph {et~al.}(1992)\citenamefont {Li},
  \citenamefont {Reimers},\ and\ \citenamefont {Dahn}}]{Li}%
  \BibitemOpen
  \bibfield  {author} {\bibinfo {author} {\bibfnamefont {W.}~\bibnamefont
  {Li}}, \bibinfo {author} {\bibfnamefont {J.~N.}\ \bibnamefont {Reimers}}, \
  and\ \bibinfo {author} {\bibfnamefont {J.~R.}\ \bibnamefont {Dahn}},\
  }\href@noop {} {\bibfield  {journal} {\bibinfo  {journal} {Phys. Rev. B:
  Condens. Matter Mater. Phys.}\ }\textbf {\bibinfo {volume} {46}},\ \bibinfo
  {pages} {3236} (\bibinfo {year} {1992})}\BibitemShut {NoStop}%
\bibitem [{\citenamefont {Guilmard}\ \emph
  {et~al.}(2003{\natexlab{a}})\citenamefont {Guilmard}, \citenamefont
  {Rougier}, \citenamefont {Gr\"{u}ne}, \citenamefont {Croguennec},\ and\
  \citenamefont {Delmas}}]{Guilmarda}%
  \BibitemOpen
  \bibfield  {author} {\bibinfo {author} {\bibfnamefont {M.}~\bibnamefont
  {Guilmard}}, \bibinfo {author} {\bibfnamefont {A.}~\bibnamefont {Rougier}},
  \bibinfo {author} {\bibfnamefont {M.}~\bibnamefont {Gr\"{u}ne}}, \bibinfo
  {author} {\bibfnamefont {L.}~\bibnamefont {Croguennec}}, \ and\ \bibinfo
  {author} {\bibfnamefont {C.}~\bibnamefont {Delmas}},\ }\href@noop {}
  {\bibfield  {journal} {\bibinfo  {journal} {J. Power Sources}\ }\textbf
  {\bibinfo {volume} {115}},\ \bibinfo {pages} {305} (\bibinfo {year}
  {2003}{\natexlab{a}})}\BibitemShut {NoStop}%
\bibitem [{\citenamefont {Guilmard}\ \emph
  {et~al.}(2003{\natexlab{b}})\citenamefont {Guilmard}, \citenamefont
  {Croguennec}, \citenamefont {Denux},\ and\ \citenamefont
  {Delmas}}]{Guilmardb}%
  \BibitemOpen
  \bibfield  {author} {\bibinfo {author} {\bibfnamefont {M.}~\bibnamefont
  {Guilmard}}, \bibinfo {author} {\bibfnamefont {L.}~\bibnamefont
  {Croguennec}}, \bibinfo {author} {\bibfnamefont {D.}~\bibnamefont {Denux}}, \
  and\ \bibinfo {author} {\bibfnamefont {C.}~\bibnamefont {Delmas}},\
  }\href@noop {} {\bibfield  {journal} {\bibinfo  {journal} {Chem. Mater.}\
  }\textbf {\bibinfo {volume} {15}},\ \bibinfo {pages} {4476} (\bibinfo {year}
  {2003}{\natexlab{b}})}\BibitemShut {NoStop}%
\bibitem [{\citenamefont {Pouillerie}\ \emph {et~al.}(2000)\citenamefont
  {Pouillerie}, \citenamefont {Croguennec}, \citenamefont {Biensan},
  \citenamefont {Willmann},\ and\ \citenamefont {Delmas}}]{Pouillerie}%
  \BibitemOpen
  \bibfield  {author} {\bibinfo {author} {\bibfnamefont {C.}~\bibnamefont
  {Pouillerie}}, \bibinfo {author} {\bibfnamefont {L.}~\bibnamefont
  {Croguennec}}, \bibinfo {author} {\bibfnamefont {P.}~\bibnamefont {Biensan}},
  \bibinfo {author} {\bibfnamefont {P.}~\bibnamefont {Willmann}}, \ and\
  \bibinfo {author} {\bibfnamefont {C.}~\bibnamefont {Delmas}},\ }\href@noop {}
  {\bibfield  {journal} {\bibinfo  {journal} {J. Electrochem. Soc.}\ }\textbf
  {\bibinfo {volume} {147}},\ \bibinfo {pages} {2061} (\bibinfo {year}
  {2000})}\BibitemShut {NoStop}%
\bibitem [{\citenamefont {Sathiyamoorthi}\ \emph {et~al.}(2007)\citenamefont
  {Sathiyamoorthi}, \citenamefont {Shakkthivel}, \citenamefont {Ramalakshmi},\
  and\ \citenamefont {Shul}}]{Sathiyamoorthi}%
  \BibitemOpen
  \bibfield  {author} {\bibinfo {author} {\bibfnamefont {R.}~\bibnamefont
  {Sathiyamoorthi}}, \bibinfo {author} {\bibfnamefont {P.}~\bibnamefont
  {Shakkthivel}}, \bibinfo {author} {\bibfnamefont {S.}~\bibnamefont
  {Ramalakshmi}}, \ and\ \bibinfo {author} {\bibfnamefont {Y.-G.}\ \bibnamefont
  {Shul}},\ }\href@noop {} {\bibfield  {journal} {\bibinfo  {journal} {J.Power
  Sources}\ }\textbf {\bibinfo {volume} {171}},\ \bibinfo {pages} {922}
  (\bibinfo {year} {2007})}\BibitemShut {NoStop}%
\bibitem [{\citenamefont {Kondo}\ \emph {et~al.}(2007)\citenamefont {Kondo},
  \citenamefont {Takeuchi}, \citenamefont {Sasaki}, \citenamefont {Kawauchi},
  \citenamefont {Itou}, \citenamefont {Hiruta}, \citenamefont {Okuda},
  \citenamefont {Yonemura}, \citenamefont {Kamiyama},\ and\ \citenamefont
  {Ukyo}}]{Kondo}%
  \BibitemOpen
  \bibfield  {author} {\bibinfo {author} {\bibfnamefont {H.}~\bibnamefont
  {Kondo}}, \bibinfo {author} {\bibfnamefont {Y.}~\bibnamefont {Takeuchi}},
  \bibinfo {author} {\bibfnamefont {T.}~\bibnamefont {Sasaki}}, \bibinfo
  {author} {\bibfnamefont {S.}~\bibnamefont {Kawauchi}}, \bibinfo {author}
  {\bibfnamefont {Y.}~\bibnamefont {Itou}}, \bibinfo {author} {\bibfnamefont
  {O.}~\bibnamefont {Hiruta}}, \bibinfo {author} {\bibfnamefont
  {C.}~\bibnamefont {Okuda}}, \bibinfo {author} {\bibfnamefont
  {M.}~\bibnamefont {Yonemura}}, \bibinfo {author} {\bibfnamefont
  {T.}~\bibnamefont {Kamiyama}}, \ and\ \bibinfo {author} {\bibfnamefont
  {Y.}~\bibnamefont {Ukyo}},\ }\href@noop {} {\bibfield  {journal} {\bibinfo
  {journal} {J. Power Sources}\ }\textbf {\bibinfo {volume} {174}},\ \bibinfo
  {pages} {1131} (\bibinfo {year} {2007})}\BibitemShut {NoStop}%
\bibitem [{\citenamefont {Cho}\ \emph {et~al.}(2000)\citenamefont {Cho},
  \citenamefont {Jung}, \citenamefont {Park}, \citenamefont {Kim},\ and\
  \citenamefont {Lim}}]{Cho}%
  \BibitemOpen
  \bibfield  {author} {\bibinfo {author} {\bibfnamefont {J.}~\bibnamefont
  {Cho}}, \bibinfo {author} {\bibfnamefont {H.}~\bibnamefont {Jung}}, \bibinfo
  {author} {\bibfnamefont {Y.}~\bibnamefont {Park}}, \bibinfo {author}
  {\bibfnamefont {G.}~\bibnamefont {Kim}}, \ and\ \bibinfo {author}
  {\bibfnamefont {H.~S.}\ \bibnamefont {Lim}},\ }\href@noop {} {\bibfield
  {journal} {\bibinfo  {journal} {J. Electrochem. Soc.}\ }\textbf {\bibinfo
  {volume} {147}},\ \bibinfo {pages} {15} (\bibinfo {year} {2000})}\BibitemShut
  {NoStop}%
\bibitem [{\citenamefont {Delmas}\ \emph {et~al.}(1993)\citenamefont {Delmas},
  \citenamefont {Saadoune},\ and\ \citenamefont {Rougier}}]{Delmas}%
  \BibitemOpen
  \bibfield  {author} {\bibinfo {author} {\bibfnamefont {C.}~\bibnamefont
  {Delmas}}, \bibinfo {author} {\bibfnamefont {I.}~\bibnamefont {Saadoune}}, \
  and\ \bibinfo {author} {\bibfnamefont {A.}~\bibnamefont {Rougier}},\
  }\href@noop {} {\bibfield  {journal} {\bibinfo  {journal} {J. Power Sources}\
  }\textbf {\bibinfo {volume} {44}},\ \bibinfo {pages} {595} (\bibinfo {year}
  {1993})}\BibitemShut {NoStop}%
\bibitem [{\citenamefont {Saadoune}\ and\ \citenamefont
  {Delmas}(1996)}]{Saadoune}%
  \BibitemOpen
  \bibfield  {author} {\bibinfo {author} {\bibfnamefont {I.}~\bibnamefont
  {Saadoune}}\ and\ \bibinfo {author} {\bibfnamefont {C.}~\bibnamefont
  {Delmas}},\ }\href@noop {} {\bibfield  {journal} {\bibinfo  {journal} {J.
  Mater. Chem.}\ }\textbf {\bibinfo {volume} {6}},\ \bibinfo {pages} {193}
  (\bibinfo {year} {1996})}\BibitemShut {NoStop}%
\bibitem [{\citenamefont {Rossen}\ \emph {et~al.}(1992)\citenamefont {Rossen},
  \citenamefont {Jones},\ and\ \citenamefont {Dahn}}]{Rossen}%
  \BibitemOpen
  \bibfield  {author} {\bibinfo {author} {\bibfnamefont {E.}~\bibnamefont
  {Rossen}}, \bibinfo {author} {\bibfnamefont {C.}~\bibnamefont {Jones}}, \
  and\ \bibinfo {author} {\bibfnamefont {J.}~\bibnamefont {Dahn}},\ }\href@noop
  {} {\bibfield  {journal} {\bibinfo  {journal} {Solid State Ionics}\ }\textbf
  {\bibinfo {volume} {57}},\ \bibinfo {pages} {311} (\bibinfo {year}
  {1992})}\BibitemShut {NoStop}%
\bibitem [{\citenamefont {Venkatraman}\ and\ \citenamefont
  {Manthiram}(2003)}]{Venkatraman}%
  \BibitemOpen
  \bibfield  {author} {\bibinfo {author} {\bibfnamefont {S.}~\bibnamefont
  {Venkatraman}}\ and\ \bibinfo {author} {\bibfnamefont {A.}~\bibnamefont
  {Manthiram}},\ }\href@noop {} {\bibfield  {journal} {\bibinfo  {journal}
  {Chem. Mater.}\ }\textbf {\bibinfo {volume} {15}},\ \bibinfo {pages} {5003}
  (\bibinfo {year} {2003})}\BibitemShut {NoStop}%
\bibitem [{\citenamefont {Mohan}\ and\ \citenamefont
  {Kalaignan}(2013)}]{Mohana}%
  \BibitemOpen
  \bibfield  {author} {\bibinfo {author} {\bibfnamefont {P.}~\bibnamefont
  {Mohan}}\ and\ \bibinfo {author} {\bibfnamefont {G.~P.}\ \bibnamefont
  {Kalaignan}},\ }\href@noop {} {\bibfield  {journal} {\bibinfo  {journal} {J.
  Electroceramics}\ }\textbf {\bibinfo {volume} {31}},\ \bibinfo {pages} {210}
  (\bibinfo {year} {2013})}\BibitemShut {NoStop}%
\bibitem [{\citenamefont {Mohan}\ and\ \citenamefont
  {Kalaignan}(2014)}]{Mohanb}%
  \BibitemOpen
  \bibfield  {author} {\bibinfo {author} {\bibfnamefont {P.}~\bibnamefont
  {Mohan}}\ and\ \bibinfo {author} {\bibfnamefont {G.~P.}\ \bibnamefont
  {Kalaignan}},\ }\href@noop {} {\bibfield  {journal} {\bibinfo  {journal} {J.
  Nanoscience and Nanotechnology}\ }\textbf {\bibinfo {volume} {14}},\ \bibinfo
  {pages} {5278} (\bibinfo {year} {2014})}\BibitemShut {NoStop}%
\bibitem [{\citenamefont {Kwon}\ \emph {et~al.}(2014)\citenamefont {Kwon},
  \citenamefont {Song},\ and\ \citenamefont {Park}}]{Kwon}%
  \BibitemOpen
  \bibfield  {author} {\bibinfo {author} {\bibfnamefont {S.~N.}\ \bibnamefont
  {Kwon}}, \bibinfo {author} {\bibfnamefont {M.~Y.}\ \bibnamefont {Song}}, \
  and\ \bibinfo {author} {\bibfnamefont {H.~R.}\ \bibnamefont {Park}},\
  }\href@noop {} {\bibfield  {journal} {\bibinfo  {journal} {Ceram. Int.}\
  }\textbf {\bibinfo {volume} {40}},\ \bibinfo {pages} {14141} (\bibinfo {year}
  {2014})}\BibitemShut {NoStop}%
\bibitem [{\citenamefont {Kim}\ \emph {et~al.}(2004)\citenamefont {Kim},
  \citenamefont {Ko}, \citenamefont {Na}, \citenamefont {Cho},\ and\
  \citenamefont {Chao}}]{Kim}%
  \BibitemOpen
  \bibfield  {author} {\bibinfo {author} {\bibfnamefont {H.-S.}\ \bibnamefont
  {Kim}}, \bibinfo {author} {\bibfnamefont {T.-K.}\ \bibnamefont {Ko}},
  \bibinfo {author} {\bibfnamefont {B.-K.}\ \bibnamefont {Na}}, \bibinfo
  {author} {\bibfnamefont {W.~I.}\ \bibnamefont {Cho}}, \ and\ \bibinfo
  {author} {\bibfnamefont {B.~W.}\ \bibnamefont {Chao}},\ }\href@noop {}
  {\bibfield  {journal} {\bibinfo  {journal} {J. Power Sources}\ }\textbf
  {\bibinfo {volume} {138}},\ \bibinfo {pages} {232} (\bibinfo {year}
  {2004})}\BibitemShut {NoStop}%
\bibitem [{\citenamefont {Yoon}\ \emph
  {et~al.}(2018{\natexlab{a}})\citenamefont {Yoon}, \citenamefont {Choi},
  \citenamefont {Jun}, \citenamefont {Zhang}, \citenamefont {Kaghazchi},
  \citenamefont {Kim},\ and\ \citenamefont {Sun}}]{Yoona}%
  \BibitemOpen
  \bibfield  {author} {\bibinfo {author} {\bibfnamefont {C.~S.}\ \bibnamefont
  {Yoon}}, \bibinfo {author} {\bibfnamefont {M.-J.}\ \bibnamefont {Choi}},
  \bibinfo {author} {\bibfnamefont {D.-W.}\ \bibnamefont {Jun}}, \bibinfo
  {author} {\bibfnamefont {Q.}~\bibnamefont {Zhang}}, \bibinfo {author}
  {\bibfnamefont {P.}~\bibnamefont {Kaghazchi}}, \bibinfo {author}
  {\bibfnamefont {K.-H.}\ \bibnamefont {Kim}}, \ and\ \bibinfo {author}
  {\bibfnamefont {Y.-K.}\ \bibnamefont {Sun}},\ }\href@noop {} {\bibfield
  {journal} {\bibinfo  {journal} {Chem. Mater.}\ }\textbf {\bibinfo {volume}
  {30}},\ \bibinfo {pages} {1808} (\bibinfo {year}
  {2018}{\natexlab{a}})}\BibitemShut {NoStop}%
\bibitem [{\citenamefont {Yoon}\ \emph
  {et~al.}(2018{\natexlab{b}})\citenamefont {Yoon}, \citenamefont {Kim},
  \citenamefont {Park}, \citenamefont {Kim}, \citenamefont {Kim}, \citenamefont
  {Kim},\ and\ \citenamefont {Sun}}]{Yoonb}%
  \BibitemOpen
  \bibfield  {author} {\bibinfo {author} {\bibfnamefont {C.~S.}\ \bibnamefont
  {Yoon}}, \bibinfo {author} {\bibfnamefont {U.-H.}\ \bibnamefont {Kim}},
  \bibinfo {author} {\bibfnamefont {G.-T.}\ \bibnamefont {Park}}, \bibinfo
  {author} {\bibfnamefont {S.~J.}\ \bibnamefont {Kim}}, \bibinfo {author}
  {\bibfnamefont {K.-H.}\ \bibnamefont {Kim}}, \bibinfo {author} {\bibfnamefont
  {J.}~\bibnamefont {Kim}}, \ and\ \bibinfo {author} {\bibfnamefont {Y.-K.}\
  \bibnamefont {Sun}},\ }\href@noop {} {\bibfield  {journal} {\bibinfo
  {journal} {ACS Energy Lett.}\ }\textbf {\bibinfo {volume} {3}},\ \bibinfo
  {pages} {1634} (\bibinfo {year} {2018}{\natexlab{b}})}\BibitemShut {NoStop}%
\bibitem [{\citenamefont {Kim}\ \emph {et~al.}(2018)\citenamefont {Kim},
  \citenamefont {Choi}, \citenamefont {Doo}, \citenamefont {Lim}, \citenamefont
  {Kim},\ and\ \citenamefont {Lee}}]{Kimb}%
  \BibitemOpen
  \bibfield  {author} {\bibinfo {author} {\bibfnamefont {H.}~\bibnamefont
  {Kim}}, \bibinfo {author} {\bibfnamefont {A.}~\bibnamefont {Choi}}, \bibinfo
  {author} {\bibfnamefont {S.~W.}\ \bibnamefont {Doo}}, \bibinfo {author}
  {\bibfnamefont {J.}~\bibnamefont {Lim}}, \bibinfo {author} {\bibfnamefont
  {Y.}~\bibnamefont {Kim}}, \ and\ \bibinfo {author} {\bibfnamefont {K.~T.}\
  \bibnamefont {Lee}},\ }\href@noop {} {\bibfield  {journal} {\bibinfo
  {journal} {J. Electrochem. Soc.}\ }\textbf {\bibinfo {volume} {165}},\
  \bibinfo {pages} {A201} (\bibinfo {year} {2018})}\BibitemShut {NoStop}%
\bibitem [{\citenamefont {Amatucci}\ \emph {et~al.}(1996)\citenamefont
  {Amatucci}, \citenamefont {Tarascon},\ and\ \citenamefont
  {Klein}}]{Amatucci}%
  \BibitemOpen
  \bibfield  {author} {\bibinfo {author} {\bibfnamefont {G.~G.}\ \bibnamefont
  {Amatucci}}, \bibinfo {author} {\bibfnamefont {J.~M.}\ \bibnamefont
  {Tarascon}}, \ and\ \bibinfo {author} {\bibfnamefont {L.~C.}\ \bibnamefont
  {Klein}},\ }\href@noop {} {\bibfield  {journal} {\bibinfo  {journal} {J.
  Electrochem. Soc.}\ }\textbf {\bibinfo {volume} {143}},\ \bibinfo {pages}
  {1114} (\bibinfo {year} {1996})}\BibitemShut {NoStop}%
\bibitem [{\citenamefont {Nishijima}\ \emph {et~al.}(2014)\citenamefont
  {Nishijima}, \citenamefont {Ootani}, \citenamefont {Kamimura}, \citenamefont
  {Sueki}, \citenamefont {Esaki}, \citenamefont {Murai}, \citenamefont
  {Fujita}, \citenamefont {Tanaka}, \citenamefont {Ohira}, \citenamefont
  {Koyama},\ and\ \citenamefont {Tanaka}}]{Nishijima}%
  \BibitemOpen
  \bibfield  {author} {\bibinfo {author} {\bibfnamefont {M.}~\bibnamefont
  {Nishijima}}, \bibinfo {author} {\bibfnamefont {T.}~\bibnamefont {Ootani}},
  \bibinfo {author} {\bibfnamefont {Y.}~\bibnamefont {Kamimura}}, \bibinfo
  {author} {\bibfnamefont {T.}~\bibnamefont {Sueki}}, \bibinfo {author}
  {\bibfnamefont {S.}~\bibnamefont {Esaki}}, \bibinfo {author} {\bibfnamefont
  {S.}~\bibnamefont {Murai}}, \bibinfo {author} {\bibfnamefont
  {K.}~\bibnamefont {Fujita}}, \bibinfo {author} {\bibfnamefont
  {K.}~\bibnamefont {Tanaka}}, \bibinfo {author} {\bibfnamefont
  {K.}~\bibnamefont {Ohira}}, \bibinfo {author} {\bibfnamefont
  {Y.}~\bibnamefont {Koyama}}, \ and\ \bibinfo {author} {\bibfnamefont
  {I.}~\bibnamefont {Tanaka}},\ }\href@noop {} {\bibfield  {journal} {\bibinfo
  {journal} {Nat. Comm.}\ }\textbf {\bibinfo {volume} {5}},\ \bibinfo {pages}
  {4553} (\bibinfo {year} {2014})}\BibitemShut {NoStop}%
\bibitem [{\citenamefont {Kresse}\ and\ \citenamefont
  {Furthm\"{u}ller}(1996{\natexlab{a}})}]{Kressea}%
  \BibitemOpen
  \bibfield  {author} {\bibinfo {author} {\bibfnamefont {G.}~\bibnamefont
  {Kresse}}\ and\ \bibinfo {author} {\bibfnamefont {J.}~\bibnamefont
  {Furthm\"{u}ller}},\ }\href@noop {} {\bibfield  {journal} {\bibinfo
  {journal} {Phys. Rev. B: Condens. Matter Mater. Phys.}\ }\textbf {\bibinfo
  {volume} {54}},\ \bibinfo {pages} {11169} (\bibinfo {year}
  {1996}{\natexlab{a}})}\BibitemShut {NoStop}%
\bibitem [{\citenamefont {Kresse}\ and\ \citenamefont
  {Furthm\"{u}ller}(1996{\natexlab{b}})}]{Kresseb}%
  \BibitemOpen
  \bibfield  {author} {\bibinfo {author} {\bibfnamefont {G.}~\bibnamefont
  {Kresse}}\ and\ \bibinfo {author} {\bibfnamefont {J.}~\bibnamefont
  {Furthm\"{u}ller}},\ }\href@noop {} {\bibfield  {journal} {\bibinfo
  {journal} {Comput. Mater. Sci.}\ }\textbf {\bibinfo {volume} {6}},\ \bibinfo
  {pages} {15} (\bibinfo {year} {1996}{\natexlab{b}})}\BibitemShut {NoStop}%
\bibitem [{\citenamefont {Klime\v{s}}\ \emph {et~al.}(2011)\citenamefont
  {Klime\v{s}}, \citenamefont {Bowler},\ and\ \citenamefont
  {Michaelides}}]{Klimesa}%
  \BibitemOpen
  \bibfield  {author} {\bibinfo {author} {\bibfnamefont {J.}~\bibnamefont
  {Klime\v{s}}}, \bibinfo {author} {\bibfnamefont {D.~R.}\ \bibnamefont
  {Bowler}}, \ and\ \bibinfo {author} {\bibfnamefont {A.}~\bibnamefont
  {Michaelides}},\ }\href@noop {} {\bibfield  {journal} {\bibinfo  {journal}
  {Phys. Rev. B: Condens. Matter Mater. Phys.}\ }\textbf {\bibinfo {volume}
  {83}},\ \bibinfo {pages} {195131} (\bibinfo {year} {2011})}\BibitemShut
  {NoStop}%
\bibitem [{\citenamefont {Klime\v{s}}\ \emph {et~al.}(2010)\citenamefont
  {Klime\v{s}}, \citenamefont {Bowler},\ and\ \citenamefont
  {Michaelides}}]{Klimesb}%
  \BibitemOpen
  \bibfield  {author} {\bibinfo {author} {\bibfnamefont {J.}~\bibnamefont
  {Klime\v{s}}}, \bibinfo {author} {\bibfnamefont {D.~R.}\ \bibnamefont
  {Bowler}}, \ and\ \bibinfo {author} {\bibfnamefont {A.}~\bibnamefont
  {Michaelides}},\ }\href@noop {} {\bibfield  {journal} {\bibinfo  {journal}
  {J. Phys.: Cond. Matt.}\ }\textbf {\bibinfo {volume} {22}},\ \bibinfo {pages}
  {022201} (\bibinfo {year} {2010})}\BibitemShut {NoStop}%
\bibitem [{\citenamefont {Perdew}\ \emph {et~al.}(1996)\citenamefont {Perdew},
  \citenamefont {Burke},\ and\ \citenamefont {Ernzerhof}}]{Predew1}%
  \BibitemOpen
  \bibfield  {author} {\bibinfo {author} {\bibfnamefont {J.~P.}\ \bibnamefont
  {Perdew}}, \bibinfo {author} {\bibfnamefont {K.}~\bibnamefont {Burke}}, \
  and\ \bibinfo {author} {\bibfnamefont {M.}~\bibnamefont {Ernzerhof}},\
  }\href@noop {} {\bibfield  {journal} {\bibinfo  {journal} {Phys. Rev. Lett.}\
  }\textbf {\bibinfo {volume} {77}},\ \bibinfo {pages} {3865} (\bibinfo {year}
  {1996})}\BibitemShut {NoStop}%
\bibitem [{\citenamefont {Grimme}\ \emph {et~al.}(2010)\citenamefont {Grimme},
  \citenamefont {Antony}, \citenamefont {Ehrlich},\ and\ \citenamefont
  {Krieg}}]{Grimme}%
  \BibitemOpen
  \bibfield  {author} {\bibinfo {author} {\bibfnamefont {S.}~\bibnamefont
  {Grimme}}, \bibinfo {author} {\bibfnamefont {J.}~\bibnamefont {Antony}},
  \bibinfo {author} {\bibfnamefont {S.}~\bibnamefont {Ehrlich}}, \ and\
  \bibinfo {author} {\bibfnamefont {H.}~\bibnamefont {Krieg}},\ }\href@noop {}
  {\bibfield  {journal} {\bibinfo  {journal} {J. Chem. Phys.}\ }\textbf
  {\bibinfo {volume} {132}},\ \bibinfo {pages} {154104} (\bibinfo {year}
  {2010})}\BibitemShut {NoStop}%
\bibitem [{\citenamefont {Chakraborty}\ \emph {et~al.}(2018)\citenamefont
  {Chakraborty}, \citenamefont {Dixit},\ and\ \citenamefont
  {Major}}]{Chakraborty}%
  \BibitemOpen
  \bibfield  {author} {\bibinfo {author} {\bibfnamefont {A.}~\bibnamefont
  {Chakraborty}}, \bibinfo {author} {\bibfnamefont {M.}~\bibnamefont {Dixit}},
  \ and\ \bibinfo {author} {\bibfnamefont {D.~T.}\ \bibnamefont {Major}},\
  }\href@noop {} {\bibfield  {journal} {\bibinfo  {journal} {npj Comput.
  Mater.}\ }\textbf {\bibinfo {volume} {4}},\ \bibinfo {pages} {60} (\bibinfo
  {year} {2018})}\BibitemShut {NoStop}%
\bibitem [{\citenamefont {Aykol}\ \emph {et~al.}(2015)\citenamefont {Aykol},
  \citenamefont {Kim},\ and\ \citenamefont {Wolverton}}]{Aykol}%
  \BibitemOpen
  \bibfield  {author} {\bibinfo {author} {\bibfnamefont {M.}~\bibnamefont
  {Aykol}}, \bibinfo {author} {\bibfnamefont {S.}~\bibnamefont {Kim}}, \ and\
  \bibinfo {author} {\bibfnamefont {C.}~\bibnamefont {Wolverton}},\ }\href@noop
  {} {\bibfield  {journal} {\bibinfo  {journal} {J. Phys. Chem. C}\ }\textbf
  {\bibinfo {volume} {119}},\ \bibinfo {pages} {19053} (\bibinfo {year}
  {2015})}\BibitemShut {NoStop}%
\bibitem [{\citenamefont {Tamura}\ \emph {et~al.}(2017)\citenamefont {Tamura},
  \citenamefont {Takai}, \citenamefont {Yabutsuka},\ and\ \citenamefont
  {Yao}}]{Tamura}%
  \BibitemOpen
  \bibfield  {author} {\bibinfo {author} {\bibfnamefont {A.}~\bibnamefont
  {Tamura}}, \bibinfo {author} {\bibfnamefont {S.}~\bibnamefont {Takai}},
  \bibinfo {author} {\bibfnamefont {T.}~\bibnamefont {Yabutsuka}}, \ and\
  \bibinfo {author} {\bibfnamefont {T.}~\bibnamefont {Yao}},\ }\href@noop {}
  {\bibfield  {journal} {\bibinfo  {journal} {J. Electrochem. Soc.}\ }\textbf
  {\bibinfo {volume} {164}},\ \bibinfo {pages} {A1514} (\bibinfo {year}
  {2017})}\BibitemShut {NoStop}%
\bibitem [{\citenamefont {Zhou}\ \emph {et~al.}(2004)\citenamefont {Zhou},
  \citenamefont {Cococcioni}, \citenamefont {Marianetti}, \citenamefont
  {Morgan},\ and\ \citenamefont {Ceder}}]{Zhou}%
  \BibitemOpen
  \bibfield  {author} {\bibinfo {author} {\bibfnamefont {F.}~\bibnamefont
  {Zhou}}, \bibinfo {author} {\bibfnamefont {M.}~\bibnamefont {Cococcioni}},
  \bibinfo {author} {\bibfnamefont {C.~A.}\ \bibnamefont {Marianetti}},
  \bibinfo {author} {\bibfnamefont {D.}~\bibnamefont {Morgan}}, \ and\ \bibinfo
  {author} {\bibfnamefont {G.}~\bibnamefont {Ceder}},\ }\href@noop {}
  {\bibfield  {journal} {\bibinfo  {journal} {Phys. Rev. B: Condens. Matter
  Mater. Phys.}\ }\textbf {\bibinfo {volume} {70}},\ \bibinfo {pages} {235121}
  (\bibinfo {year} {2004})}\BibitemShut {NoStop}%
\bibitem [{\citenamefont {Ouma}\ \emph {et~al.}(2012)\citenamefont {Ouma},
  \citenamefont {Mapelu}, \citenamefont {Makau}, \citenamefont {Amolo},\ and\
  \citenamefont {Maezono}}]{2012OUM}%
  \BibitemOpen
  \bibfield  {author} {\bibinfo {author} {\bibfnamefont {C.~N.~M.}\
  \bibnamefont {Ouma}}, \bibinfo {author} {\bibfnamefont {M.~Z.}\ \bibnamefont
  {Mapelu}}, \bibinfo {author} {\bibfnamefont {N.~W.}\ \bibnamefont {Makau}},
  \bibinfo {author} {\bibfnamefont {G.~O.}\ \bibnamefont {Amolo}}, \ and\
  \bibinfo {author} {\bibfnamefont {R.}~\bibnamefont {Maezono}},\ }\href@noop
  {} {\bibfield  {journal} {\bibinfo  {journal} {Phys. Rev. B: Condens. Matter
  Mater. Phys.}\ }\textbf {\bibinfo {volume} {86}},\ \bibinfo {pages} {104115}
  (\bibinfo {year} {2012})}\BibitemShut {NoStop}%
\bibitem [{\citenamefont {Kittel}(2004)}]{Kittel}%
  \BibitemOpen
  \bibfield  {author} {\bibinfo {author} {\bibfnamefont {C.}~\bibnamefont
  {Kittel}},\ }\href@noop {} {\emph {\bibinfo {title} {InIntroduction to Solid
  State Physics}}},\ \bibinfo {edition} {8th}\ ed.\ (\bibinfo  {publisher}
  {John Wiley \& Sons, Inc.},\ \bibinfo {address} {New York},\ \bibinfo {year}
  {2004})\BibitemShut {NoStop}%
\bibitem [{\citenamefont {Hongo}\ and\ \citenamefont {Maezono}(2017)}]{Hongo}%
  \BibitemOpen
  \bibfield  {author} {\bibinfo {author} {\bibfnamefont {K.}~\bibnamefont
  {Hongo}}\ and\ \bibinfo {author} {\bibfnamefont {R.}~\bibnamefont
  {Maezono}},\ }\href@noop {} {\bibfield  {journal} {\bibinfo  {journal} {J.
  Chem. Theory Comput.}\ }\textbf {\bibinfo {volume} {13}},\ \bibinfo {pages}
  {5217} (\bibinfo {year} {2017})}\BibitemShut {NoStop}%
\bibitem [{\citenamefont {Yao}\ and\ \citenamefont {Kanai}(2017)}]{Yao}%
  \BibitemOpen
  \bibfield  {author} {\bibinfo {author} {\bibfnamefont {Y.}~\bibnamefont
  {Yao}}\ and\ \bibinfo {author} {\bibfnamefont {Y.}~\bibnamefont {Kanai}},\
  }\href@noop {} {\bibfield  {journal} {\bibinfo  {journal} {J. Chem. Phys.}\
  }\textbf {\bibinfo {volume} {146}},\ \bibinfo {pages} {224105} (\bibinfo
  {year} {2017})}\BibitemShut {NoStop}%
\bibitem [{\citenamefont {Schipper}\ \emph {et~al.}(2017)\citenamefont
  {Schipper}, \citenamefont {Erickson}, \citenamefont {Erk}, \citenamefont
  {Shin}, \citenamefont {Chesneau},\ and\ \citenamefont
  {Aurbach}}]{Schipper01012017}%
  \BibitemOpen
  \bibfield  {author} {\bibinfo {author} {\bibfnamefont {F.}~\bibnamefont
  {Schipper}}, \bibinfo {author} {\bibfnamefont {E.~M.}\ \bibnamefont
  {Erickson}}, \bibinfo {author} {\bibfnamefont {C.}~\bibnamefont {Erk}},
  \bibinfo {author} {\bibfnamefont {J.-Y.}\ \bibnamefont {Shin}}, \bibinfo
  {author} {\bibfnamefont {F.~F.}\ \bibnamefont {Chesneau}}, \ and\ \bibinfo
  {author} {\bibfnamefont {D.}~\bibnamefont {Aurbach}},\ }\href@noop {}
  {\bibfield  {journal} {\bibinfo  {journal} {J. Electrochem. Soc.}\ }\textbf
  {\bibinfo {volume} {164}},\ \bibinfo {pages} {A6220} (\bibinfo {year}
  {2017})}\BibitemShut {NoStop}%
\bibitem [{\citenamefont {Anisimov}\ \emph {et~al.}(1991)\citenamefont
  {Anisimov}, \citenamefont {Zaanen},\ and\ \citenamefont
  {Andersen}}]{PhysRevB.44.943}%
  \BibitemOpen
  \bibfield  {author} {\bibinfo {author} {\bibfnamefont {V.~I.}\ \bibnamefont
  {Anisimov}}, \bibinfo {author} {\bibfnamefont {J.}~\bibnamefont {Zaanen}}, \
  and\ \bibinfo {author} {\bibfnamefont {O.~K.}\ \bibnamefont {Andersen}},\
  }\href@noop {} {\bibfield  {journal} {\bibinfo  {journal} {Phys. Rev. B:
  Condens. Matter Mater. Phys.}\ }\textbf {\bibinfo {volume} {44}},\ \bibinfo
  {pages} {943} (\bibinfo {year} {1991})}\BibitemShut {NoStop}%
\bibitem [{\citenamefont {Ong}\ \emph {et~al.}(2013)\citenamefont {Ong},
  \citenamefont {Richards}, \citenamefont {Jain}, \citenamefont {Hautier},
  \citenamefont {Kocher}, \citenamefont {Cholia}, \citenamefont {Gunter},
  \citenamefont {Chevrier}, \citenamefont {Persson},\ and\ \citenamefont
  {Ceder}}]{Ong}%
  \BibitemOpen
  \bibfield  {author} {\bibinfo {author} {\bibfnamefont {S.~P.}\ \bibnamefont
  {Ong}}, \bibinfo {author} {\bibfnamefont {W.~D.}\ \bibnamefont {Richards}},
  \bibinfo {author} {\bibfnamefont {A.}~\bibnamefont {Jain}}, \bibinfo {author}
  {\bibfnamefont {G.}~\bibnamefont {Hautier}}, \bibinfo {author} {\bibfnamefont
  {M.}~\bibnamefont {Kocher}}, \bibinfo {author} {\bibfnamefont
  {S.}~\bibnamefont {Cholia}}, \bibinfo {author} {\bibfnamefont
  {D.}~\bibnamefont {Gunter}}, \bibinfo {author} {\bibfnamefont {V.~L.}\
  \bibnamefont {Chevrier}}, \bibinfo {author} {\bibfnamefont {K.~A.}\
  \bibnamefont {Persson}}, \ and\ \bibinfo {author} {\bibfnamefont
  {G.}~\bibnamefont {Ceder}},\ }\href@noop {} {\bibfield  {journal} {\bibinfo
  {journal} {Comput. Mater. Sci.}\ }\textbf {\bibinfo {volume} {68}},\ \bibinfo
  {pages} {314} (\bibinfo {year} {2013})}\BibitemShut {NoStop}%
\bibitem [{\citenamefont {R.Tibshirani}(1996)}]{Tibshirani}%
  \BibitemOpen
  \bibfield  {author} {\bibinfo {author} {\bibnamefont {R.Tibshirani}},\
  }\href@noop {} {\bibfield  {journal} {\bibinfo  {journal} {J. R. Stat. Soc.
  Ser.}\ }\textbf {\bibinfo {volume} {58}},\ \bibinfo {pages} {267} (\bibinfo
  {year} {1996})}\BibitemShut {NoStop}%
\end{thebibliography}%

\end{document}